# Thermo Activated Hysteresis on High Quality Graphene/*h*-BN Devices

A. R. Cadore,[1,a] E. Mania,[1] K. Watanabe,[2] T. Taniguchi,[2] R. G. Lacerda,[1] and L. C. Campos[1,a]

[1]*Departamento de Física, Universidade Federal de Minas Gerais, Belo Horizonte, 30123-970, Brazil*

[2]*Advanced Materials Laboratory, National Institute for Materials Science, 1-1Namiki, 305-0044, Japan*

[a] *Electronic mail: alissoncadore@gmail.com; lccampos@fisica.ufmg.br*

We report on gate hysteresis in resistance on high quality graphene/*h*-BN devices. We observe a thermal activated hysteretic behavior in resistance as a function of the applied gate voltage at temperatures above 375K. In order to investigate the origin of the hysteretic phenomenon, we design heterostructures involving graphene/*h*-BN devices with different underlying substrates such as: $SiO_2$/Si and graphite; where heavily doped silicon and graphite are used as a back gate electrodes, respectively. The gate hysteretic behavior of the resistance shows to be present only in devices with an *h*-BN/$SiO_2$ interface and is dependent on the orientation of the applied gate electric field and sweep rate. Finally, we suggest a phenomenological model, which captures all of our findings based on charges trapped at the *h*-BN/$SiO_2$. Certainly, such hysteretic behavior in graphene resistance represents a technological problem for the application of graphene devices at high temperatures, but conversely, it can open new routes for applications on digital electronics and graphene memory devices.

Graphene is currently attracting attention as a potential building blocks for future nanoelectronics devices due to its physical and electrical properties.[1–3] The ongoing development on graphene device technology is supported by high mobility and fast response of graphene field effect transistors (FET).[4] However, the intrinsic graphene characteristics can be significantly affected by the underlying substrates. For instance: hysteretic behaviors of graphene FET have been observed in graphene devices on different substrates,[5–13] opening several questions regarding the ideal platform for graphene transistors, limitations concerning temperature and influence of the environment on their properties. Out of a variety of dielectric materials, so far, hexagonal boron nitride (*h*-BN) stands up as the best platform for graphene devices. Up to now, graphene transistors on *h*-BN substrates show low charge inhomogeneity, and they free of hysteretic effects at temperatures ranging from room temperature (RT) down to low temperatures.[14–16] Also, a significant amount of attention regarding graphene *h*-BN heterostructures have been performed at low temperatures, demonstrating its ultrahigh mobility and showing new physical properties.[3,17] Moreover, as the demand for technological applications of graphene increases, various possibilities start to be considered, for instance, graphene devices based on high field and high current measurements.[18–20] Curiously, little attention has been given to graphene/*h*-BN devices working at operating temperatures of transistors, sensors and digital devices. Therefore, it is crucial to understand the temperature influence on electrical properties of graphene supported on *h*-BN, because graphene has great ability to tolerate the impact of temperature,[21] added to the fact that *h*-BN can improve heat dissipation.[22]

In this work, we present an investigation on the electronic properties of graphene/*h*-BN/SiO$_2$/Si (GBN/SiO$_2$) devices as a function of temperature (from 4K up to 500K). We report on advent of a thermo-activated hysteresis in the resistance of graphene devices. This hysteresis causes a distortion on the transconductance of graphene FETs, resulting in a misleading interpretation of the true values of the field effect mobility and minimum of conductivity. This phenomenon promotes changes in the graphene density of charges, depending on the orientation of the applied electric field. Interestingly, in devices that we incorporate a graphite layer in between *h*-BN and SiO$_2$ the hysteretic behavior is totally suppressed. We nominate GBN/graphite devices a heterostructure composed by graphene/*h*-BN/graphite/SiO$_2$/Si. Our experiments show that the hysteretic effect on GBN/SiO$_2$ devices originates from trapped charges at the interface in between *h*-BN and SiO$_2$. Besides, it shows that there are still important and open issues to be addressed in order to fully understand graphene/*h*-BN heterostructures. These problems need to be solved in order to employ high quality graphene devices on electronic applications.

Here, we investigate three different graphene heterostructures, all prepared by standard scotch tape method; i) graphene device on SiO$_2$/Si substrate (G/SiO$_2$) with 285nm of SiO$_2$ grown on top of highly p-type doped Si wafers; ii) graphene devices atop of *h*-BN flakes,[23] which are previously exfoliated to SiO$_2$/Si substrate (see Fig. 1(a)). In these architectures, we use the doped Si substrate as a back gate bias; iii) graphene devices atop of *h*-BN flakes previously transferred atop of a ~20nm thick graphite (see Fig. 1(b)). In this architecture, we use the graphite layer as a local back gate terminal excluding the *h*-BN/SiO$_2$ interface. After each material transfer, samples are submitted to thermal annealing at 623K with constant flow of Ar:H$_2$ (300:700sccm) for 3.5h to remove organic residues. The *h*-BN and graphene flakes are characterized by atomic force microscope (AFM) to check flatness and cleanness, and by optical analysis and Raman spectroscopy to identify monolayer graphene. Finally, metallic contacts Cr/Au (1/50nm) are patterned by standard electron-beam lithography and by thermal metal deposition. Afterwards, graphene is shaped into the desired geometry by oxygen plasma etching. To remove polymer residues reminiscent from the lithography processes, devices are submitted to a new thermal annealing. Fig. 1(c) shows the AFM image of an ordinary GBN/SiO$_2$ device. The figure depicts that the samples are free of bubble, wrinkles or polymer residues on graphene region, indicating the GBN/SiO$_2$ device has an ultraclean graphene surface.

To investigate the role of the *h*-BN/SiO$_2$ interface, we examine four different samples in the GBN/SiO$_2$ architecture and three samples in the GBN/graphite configuration. In order to attain the best surface flatness, we prepare devices with *h*-

BN thickness in between 15nm to 22nm.[17] We perform electrical transport measurements in linear four-terminal and Hall bar geometry using standard lock-in techniques at a frequency of 17Hz with a current bias of 100nA. We carry out all the measurements in a Janis ST-300SH cryostat system that allows *in situ* measurements at temperatures from 4K up to 500K in vacuum at ~ $1\times10^{-6}$ Torr. For gate hysteresis measurements, we sweep the back gate bias ($V_G$) in the sequence $0\rightarrow -V_0$, $-V_0\rightarrow +V_0$, and $V_0\rightarrow 0$; where $V_0$ is the maximum applied gate bias. In the inverse sequence, there is no significant change in the measurements. We repeat each hysteresis loop several times without any changes to ascertain the reproducibility of the measurements. Finally, it is important to mention that for all GBN/SiO$_2$ and GBN/graphite samples the initial measurements at RT exhibit the position of the charge neutrality point (CNP) close to $V_G$=0V, as shown in the inset of Fig. 1(d). From this inset, one can see that there is a minor doping, and the GBN/SiO$_2$ sample do not show any hysteresis in resistance at 300K. For the G/SiO$_2$ samples at RT, the CNP indicated a p-type doping must probably due to adsorbed water on the graphene surface.[5] However, after a standard overnight conditioning at $T$=500K, the G/SiO$_2$ device shows an n-type doping, normally related to well-known activation of dangling bonds from SiO$_2$.[24]

We now present our main experimental observations. Fig. 1(d) shows a pronounced hysteresis in the resistance (*R*) when we sweep the gate voltage at $T$=500K in a GBN/SiO$_2$ sample. The blue curve in Fig. 1(d) depicts the measurement during the forward sweep (-40V$\rightarrow$ +40V) of $V_G$, whereas the red dashed curve shows the backward sweeping (+40V$\rightarrow$ -40V); in both cases the sweep rate is 0.17V/s. As one can see, there is a considerable difference between the two curves both in shape, and position of the CNP. The CNP shifts positively (negatively) for forward (backward) gate bias, being positioned at +4V (-18V), in relation to zero voltage. Also, these positions remains fixed even for several sweep loops, maintaining a constant difference between both CNP ($\Delta V$=22V). The results showed here were obtained for sample with a 22nm thick *h*-BN. However, a device with 15nm thick *h*-BN depicts a similar behavior, which indicates that the *h*-BN thickness does not contribute to the hysteresis effect. Likewise, we believe that the hysteretic phenomenon is not caused by water or other molecules adsorbed on graphene surface.[8] The hysteresis in the resistance remains unchanged even after keeping the sample for two weeks or longer at $T$=500K and vacuum. Also, from the relative shift of the gate voltage sweep, it is possible to determine the direction of the hysteresis: when the CNP shifts positively (negatively) in relation to $V_G$=0V for negative (positive) gate voltages, the sample shows negative (positive) hysteresis.[5] Consequently our GBN/SiO$_2$ samples show a negative hysteresis. This negative behavior is normally attributed to capacitive coupling to graphene device, and it could be caused by a polarization of water molecules trapped in between graphene/*h*-BN.[5,15]

To further elucidate the origin of the hysteretic effect, heterostructures with a graphite layer in between GBN and SiO$_2$ (GBN/graphite) were fabricated. Whereas, we observe a clear hysteretic behavior in GBN/SiO$_2$ samples, we do not see any hysteresis in resistance of GBN/graphite or G/SiO$_2$ devices. Fig. 1(e) depicts the $R$ x $V_G$ for the GBN/graphite device sweeping the gate voltage (at $T$=500K) from -2.5V to +2.5V. For both sweep gate directions, one can observe the position of the CNP close to $V_G$= -0.05V (with sweep rate of 0.025V/s). It is important to mention that in both GBN/SiO$_2$ and GBN/graphite devices the maximum transversal applied electric field is about 0.12V/nm. In order to compare both results, we also measured a graphene directly to SiO$_2$ – see inset Fig. 1(e). The G/SiO$_2$ device shows a small hysteresis effect, indicating that the contribution to hysteresis from charge trapped only on the SiO$_2$ surface or possibly water between graphene and SiO$_2$ is small ($\Delta V$~1.5V). From these reasons, one can argue that the origin of the observation of the hysteresis in the graphene resistance is related to the $h$-BN/SiO$_2$ interface. Other possible hysteretic sources such as: water between graphene and $h$-BN, or surface/bulk defects in the $h$-BN can be ignored since there is no hysteresis effect on the GBN/graphite samples.

Fig. 2(a) depicts the position of the CNP as the gate bias sweeps in a loop from -40V to +40V (forward/backwards directions) with a fixed sweep rate of 0.17V/s as a function of temperature (4K–500K). Interestingly, at temperatures below RT (down to 4K) GBN/SiO$_2$ measurements do not show any hysteretic effect in the graphene resistance. This indicates that charge traps get frozen out at lower temperatures, suppressing the gate hysteresis.[14,15,25] In addition, at temperature of 4K charge carrier mobility is about 40.000cm$^2$/Vs, asserting the high quality of the device. As seen in Fig. 2(a), at temperature $T$=375K the hysteresis starts to take place and a split between the CNPs appears in a sweep loop. Moreover, as the temperature gets higher, a more pronounced splitting between both CNPs points is observed, indicating that the hysteresis effect in GBN/SiO$_2$ system seems to be thermo-activated. Fig. 2(b) shows the GBN/graphite measurements with a gate bias loop from -2.5V to +2.5V, with a fixed sweep rate of 0.025V/s. From this figure, one can see that there is no splitting between the two CNP, demonstrating again that the hysteretic behavior is absent in devices free of $h$-BN/SiO$_2$ interface.

Now we investigate the hysteresis in the resistance with respect to the sweeping rate. We change the sweeping rate from 0.050V/s to 2.7V/s at a fixed temperature of $T$=500K – temperature which we observe the most pronounced hysteric effect. Fig. 2(c) depicts the position of both CNP against the sweep rate for one GBN/SiO$_2$ sample with a 15nm thick $h$-BN. The hysteretic behavior turns out to be larger as the rate increases, and the splitting between the CNPs seems to reach a saturation around 2V/s, with a difference between both CNPs peaks in order of $\Delta V$~45V. While the hysteresis in the resistance strongly depends on the sweeping rate, it seems to be independent of the maximum gate voltage. Therefore, our results show that charges are not injected into the $h$-BN/SiO$_2$ interface by the applied electric field.[5] Another intriguing aspect of this hysteretic effect is the time memory related with the shifts of the CNP. In a forward sweep, the resistance is stationary unless the sweep across the CNP. However, in a backward measurement, if the CNP is not reached by the gate sweep, it takes around $t$=15min for the system to restore to the forward condition resistance at $T$=500K, and more than 4h at $T$=375K.

One possibility to explain the hysteresis in resistance of the GBN/SiO$_2$ devices would be a charging effect of the interface between both dielectrics. For instance, bulk and surface defects in the h-BN were already proposed in order to explain photoinduced doping in the graphene channel.[25] Thus, since SiO$_2$ generally contains a substantial amount of defects,[11,24] the hysteretic behavior can be phenomenologically understood in terms of charge-trap memory operation, where the h-BN/SiO$_2$ interface defect may trap or detrap charges when the total electric field ($\vec{E}$) flips up and down. When the gate bias is applied, an electric field is induced at the heterostructure. Then, the h-BN/SiO$_2$ interface pins an amount of charge that is able to add an extra amount of charge to graphene channel as illustrated in the Fig. 3(a). However, once the gate bias exceeds the CNP ($V_G^{CNP}$) electric field flips upside-down and we observe a shift of the CNP. This is equivalent of an inversion of the trapped charges at the interface as we propose in Fig. 3(b). The changes in the CNP position with the sweep rate can be associated with mobile charges trapped to the interface.[11] These charges move easily at the h-BN/SiO$_2$ interface during the gate sweep, changing the graphene resistance and the total electric field. Both defects show to be not significant at RT down to 4K, but play an important role at high temperature. Also, the temperature dependence can be explained in terms of thermal energy. It is necessary to add some minimum energy to activate those dielectrics defects, or change states at the h-BN/SiO$_2$ interface. However, the origin of those defects is still unknown

For graphene devices with hysteretic phenomenon, the charge carrier density cannot be directly related to the gate electrostatic potential via a simple capacitor model anymore. The density of charge in the graphene will now be dependent on the influence of the gate potential and of an unknown potential that is responsible for the hysteresis.[5,7,15] Even though, we are not aware of the origin of the trapped charges, their amount is estimated by the split between the CNP ($\Delta V$) using $N_{trap} = C_{ox} \cdot \Delta V/q$, where $C_{ox}$ is the capacitance of the dielectrics (h-BN and SiO$_2$), and $q$ is electron charge. Fig. 3(c) shows the effective charge trapped as a function of temperature. As one can see, as the temperature increases ($T>375K$), more charges are trapped in the interface, resulting to an enhancement of the hysteretic effect in the graphene resistance. Besides, a careful analysis of the curves showed in Fig. 1(d) shows that as soon $V_G$ changes during the sweep, both curves get distorted, and the intrinsic properties of graphene device cannot be properly determined. For instance, Fig. 4(a) shows the minimum of conductivity ($\sigma_0$) in a GBN/SiO$_2$ sample from 4K to 500K. One would expect that as the temperature gets higher, the imprecision in energy to describe the value of $\sigma_0$ should increase,[26] and phonons also would start to act in the system causing a reduction of the transconductance (dI/dV).[27] However, for $T>375K$ is possible to observe a complete distortion in the position of the $\sigma_0$ (Fig. 4(a)), indicating that the trapped charges impairs considerably the graphene electronic properties, even close to the CNP. This distortion cannot be addressed only to temperature, because for a GBN/graphite sample the $\sigma_0$ follows the expected behavior (Fig. 4(b)).[26,28]

At high temperatures, graphene FET transconductance also gets distorted. Fig. 4(c) depicts that there is an increase of the transconductance, and a considerable difference between the forward and backward results. This effect is caused by the trapped charge at the $h$-BN/SiO$_2$ interface that masks the total charge density induced by gate to the graphene sheet. This distortion leads to an incorrect charge mobility analysis, resulting in a misleading interpretation of the real values of the FET mobility as the temperature gets above 375K. On the other hand, this characteristic is completely absent at samples with a graphite layer (Fig. 4(d)), demonstrating a continuous behavior from 4K up to 500K, as expected.[26–28] In fact, our results prove that the hysteresis may cause uncertainty in measuring the conductance and the analysis of the field effect mobility, which may lead to large discrepancies in the results, once the sweeping rate and direction from the gate bias is not reported.

To conclude, we have investigated the use of graphene/$h$-BN/SiO$_2$/Si devices as a function of temperature. We report a pronounced hysteresis in the resistance of the graphene devices for temperature measurements above 375K. The hysteretic phenomenon observed here shows to be thermo-activated, increasing up to 500K, and it is dependent of the gate sweep rate. The hysteresis vanishes completely in graphene/$h$-BN/graphite/SiO$_2$/Si devices when the graphite is used to eliminate the $h$-BN/SiO$_2$ interface, restoring the intrinsic graphene characteristics. It is proposed that this hysteresis behavior is addressed to site defects at both dielectrics that may enhance the charge trap density at the $h$-BN/SiO$_2$ interface. This observation evidences that there are still technological problems to be solved to use $h$-BN as substrates for commercial graphene field effect transistors. Finally, the hysteretic behavior in low disordered GBN/SiO$_2$ devices could also be used for memory devices operating at high temperatures.

**Figure 1:**

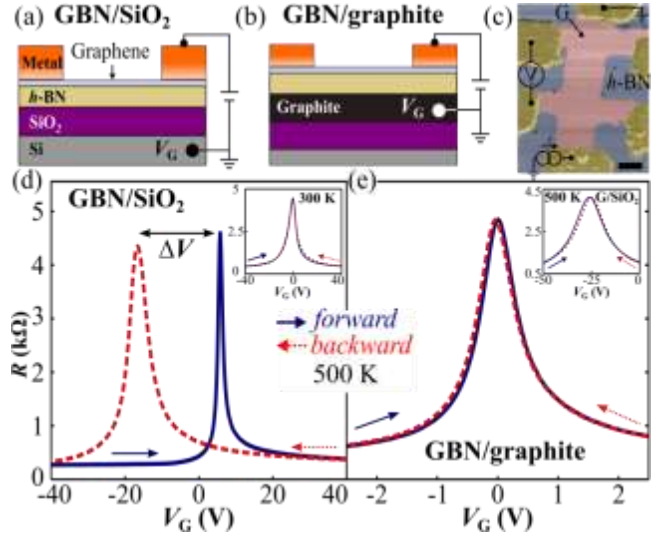

Fig. 1: Sketch of a graphene field effect device deposited on: (a) $h$-BN/SiO$_2$/Si (GBN/SiO$_2$), and; (b) $h$-BN/graphite/SiO$_2$/Si (GBN/graphite) heterostructures. (c) AFM false color image of hall bar geometry of a GBN/SiO$_2$ device. Graphene (G) is indicated in light pink in the image, while the 22nm thick $h$-BN is in light blue. The scale bar is 500nm. (d) Resistance ($R$) hysteresis recorded for a back gate bias from -40V to +40V at 500K, with a sweep rate of 0.17V/s for a GBN/SiO$_2$ device. $\Delta V$ depicts the CNP difference between the forward and backward back gate bias sweep. Inset shows the measurement at 300K for same sample, indicating minor doping and no hysteresis in the resistance. (e) $R$ x $V_G$ for a back gate bias from -2.5V to +2.5V, with a sweep rate of 0.025V/s for a GBN/graphite device with no hysteresis being observed. Inset shows the measurement for a graphene directly atop of SiO$_2$ (G/SiO$_2$) for gate bias from -50V to 0V, with a sweep rate of 0.25V/s at $T$=500K. The direction of the gate bias sweep is denoted by the arrows.

**Figure 2:**

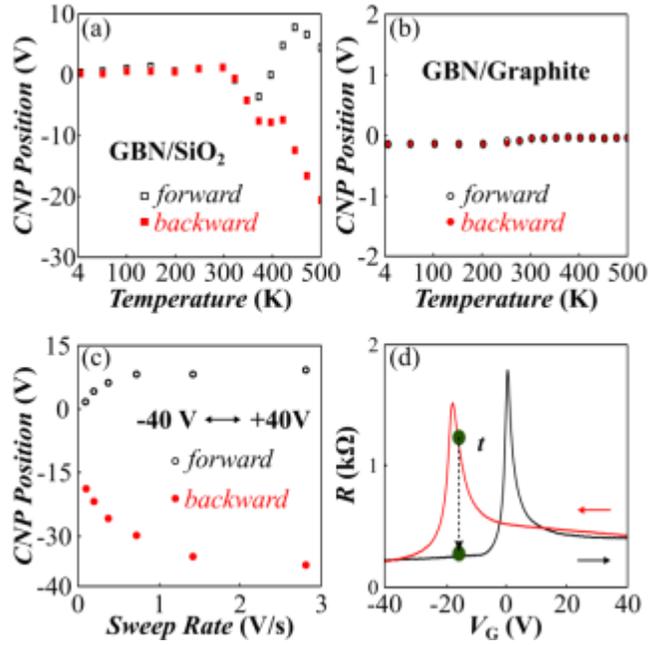

Fig. 2: (a) Position of CNP as a function of the temperature for a GBN/SiO$_2$ device for a fixed gate bias of $V_0= \pm 40$V loop, and sweep rate of 0.17V/s; (b) Position of CNP versus temperature for a GBN/graphite device for a fixed gate bias of $V_0= \pm 2.5$V loop, and sweep rate of 0.025V/s. (c) Position of CNP versus sweep rate of the back gate bias when the gate is swept from -40V to 40V at a fixed temperature $T$=500K. (d) $R$ x $V_G$ for a GBN/SiO$_2$ device for a sweep rate of 0.17V/s; The black dashed arrow in the figure represents the time ($t$) that is needed to decay from the backward resistance to the forward resistance value if the CNP is not reached by the gate bias sweep. The green balls are in a fixed $V_G = -15$V. The data shown in figure (a) is from a sample with a 22nm thick $h$-BN, while in figure (b) with a 15nm thick $h$-BN.

**Figure 3:**

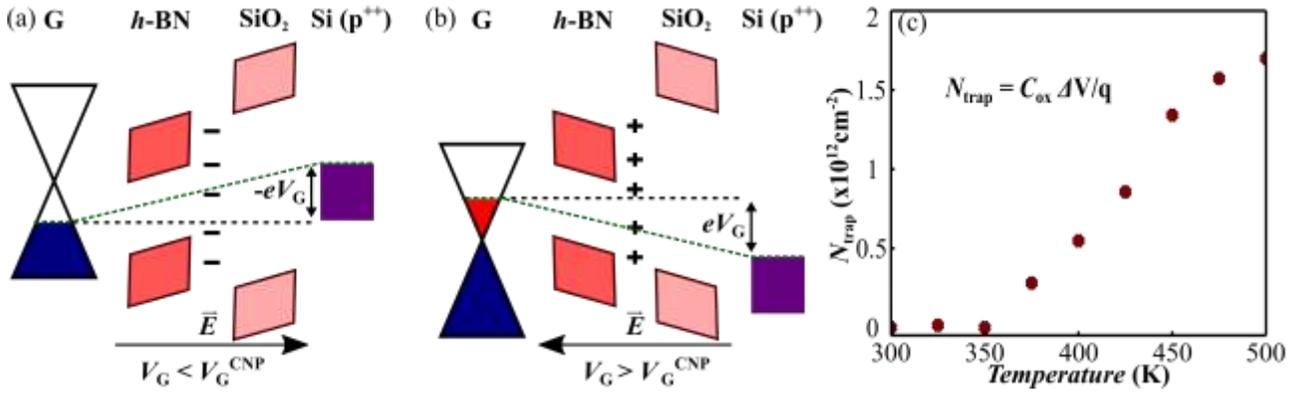

Fig. 3: Schematics of the band structure of the GBN/SiO$_2$ heterostructure and illustration of the electric field created by the gate bias ($\vec{E}$) and the inversion of the trapped charges. Figures a) and b) show the trapped charges at the $h$-BN/SiO$_2$ interface that induces electron depletion (accumulation) to graphene channel during the forward (backward) gate bias sweep in consequence of the electric field applied by the gate bias, respectively. $V_G$ is back gate bias applied, while the $V_G^{CNP}$ is voltage at the CNP. b) Effective charge trapped ($N_{trap}$) versus temperature of the same data taken in Fig. 2(a). The total amount of effective charge trapped is estimated by the split between the CNP ($\Delta V$) using $N_{trap} = C_{ox} \cdot \Delta V/q$; where $C_{ox}$ is the capacitance of the dielectrics ($h$-BN and SiO$_2$), and $q$ is electron charge.

**Figure 4:**

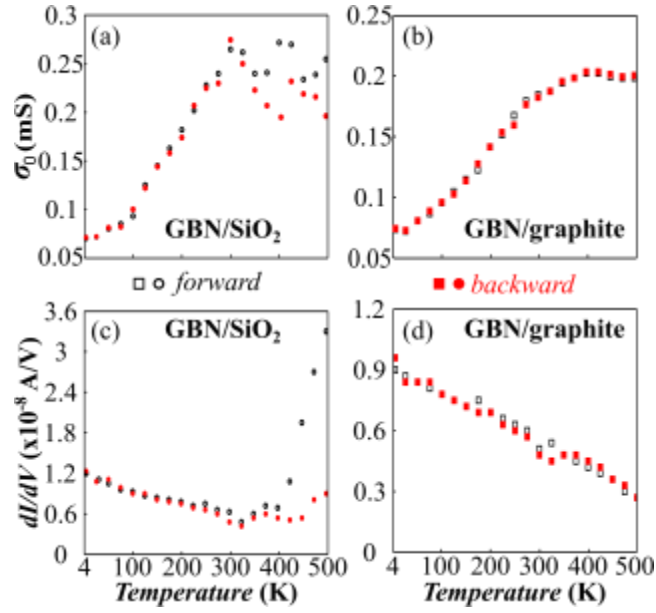

Fig. 4: (a) and (b) Minimum conductivity ($\sigma_0$) as a function of temperature for the GBN/SiO$_2$ and GBN/graphite, respectively. (c) and (d) graphene FET transconductance (dI/dV) as a function of temperature for the GBN/SiO$_2$ and GBN/graphite, respectively. The trapped charge causes uncertainty in the measurements of both conductance and transconductance of graphene devices. For the GBN/SiO$_2$ device the data is taken from -40V to +40V loop with a fixed sweep rate of 0.17V/s, while for the GBN/graphite we apply a gate bias from -2.5V to +2.5V, with a sweep rate of 0.025V/s.

ACKNOWLEDGMENT

This work was supported by CAPES, Fapemig, CNPq, Rede de Nano-Instrumentação and INCT/Nanomateriais de Carbono. The authors are thankful to Centro Brasileiro de Pesquisas Físicas (CBPF) and Centro de Componentes Semicondutores (CCS) for providing an e-beam lithography system, and Lab Nano at UFMG for allowing the use of atomic force microscopy.